\documentclass[11pt]{article}
\usepackage{graphicx}
\usepackage{dsfont}
\begin{document}
\title{Direct evidence for a Coulombic phase in monopole-suppressed SU(2) lattice gauge theory}
\author{Michael Grady\\
Department of Physics\\ State University of New York at Fredonia\\
Fredonia NY 14063 USA\\ph:(716)673-4624, fax:(716)673-3347, email: grady@fredonia.edu}
\date{\today}
\maketitle
\thispagestyle{empty}
\begin{abstract}
Further evidence is presented for the existence of a non-confining phase at weak coupling in 
SU(2) lattice gauge theory.  Using Monte Carlo simulations with the standard Wilson action, gauge-invariant 
SO(3)-Z2 monopoles, which are strong-coupling lattice artifacts,
have been  seen to undergo a percolation transition exactly at the phase transition previously seen using
Coulomb-gauge methods, with an infinite lattice critical point near $\beta = 3.2$.  The theory 
with both Z2 vortices and monopoles and
SO(3)-Z2 monopoles eliminated is simulated in the strong coupling ($\beta = 0$) limit on lattices up to $60^4$. 
Here, as in the high-$\beta$ phase of the Wilson action theory, finite size scaling shows it 
spontaneously breaks the remnant symmetry left
over after Coulomb gauge fixing. 
Such a symmetry breaking precludes the potential from having a linear term.  The monopole restriction
appears to prevent the transition to a confining phase at any $\beta$.
Direct measurement of the instantaneous Coulomb potential shows a Coulombic
form with moderately running coupling possibly approaching an infrared fixed point of $\alpha \sim 1.4$. 
The Coulomb potential is measured to 50 lattice spacings and 2 fm. 
A short-distance fit to
the 2-loop perturbative potential is used to set the scale. High precision 
at such long distances 
is made possible through the use of open boundary conditions, which was previously found to cut 
random and systematic errors of the Coulomb gauge fixing procedure dramatically.
The Coulomb potential agrees with the gauge-invariant 
interquark potential measured with smeared Wilson loops on periodic lattices as far as 
the latter can be practically 
measured with similar statistics data.

\end{abstract}


\section{Introduction}
Transforming lattice configurations to the minimal Coulomb gauge allows the definition of a local
order parameter for lattice gauge theory, the Coulomb magnetization, 
which is simply the expectation value of the three-space average
of the fourth-direction
pointing link.  This quantity is defined on spacelike hyperlayers because there is a separate SU(2) global
remnant symmetry left on each hyperlayer after Coulomb gauge fixing, so there is technically one order 
parameter per
hyperlayer.  Spontaneous breaking of this order parameter can occur and has been
seen to occur in Monte Carlo simulations of SU(2) lattice gauge theory at weak coupling\cite{coulgauge}. 
This result appears to hold also on the infinite lattice as determined by standard finite size scaling methods such
as Binder cumulant crossings and scaling collapse fits.  The infinite lattice critical point was reported as
$\beta _c = 3.18\pm 0.08$. Ref. \cite{coulgauge} also shows that this transition
is connected to the well known magnetic phase transition in the 3-d O(4) Heisenberg model, through extending the
coupling space to one in which vertical plaquettes (those with one timelike link) and horizontal plaquettes (purely
spacelike) have different couplings. Such a connection, along with the symmetry-breaking nature of the 
phase transition
makes the usually assumed non-existence of such a phase transition paradoxical.  
A proof of the existence of this phase transition on the infinite lattice, based on its connection to 
the Heisenberg model, is given in Ref.~\cite{proof}.
In the Coulomb Gauge,  where a local order parameter
can be defined, the
lattice gauge theory appears to be behaving much like its cousin spin model, having a 
ferromagnetic phase at weak coupling (analogous 
to low temperature for the magnetic analogue).

However, the existence of such a phase transition requires giving up a long-standing assumption that the
non-abelian lattice gauge theories confine in the continuum limit.  This is because it has been shown that
spontaneous breaking of the remnant gauge symmetry necessarily leads to a non-confining instantaneous Coulomb
potential. Since this potential is an upper limit to the standard interquark potential that also cannot be confining\cite{zw,mari}.
Although this means that to show non-confinement, demonstration of remnant symmetry breaking is sufficient,
it would be interesting to see what the potential actually looks like in the weak-coupling phase, especially outside the
perturbative region.  In particular it would be very interesting to see whether the running coupling continues
to increase or approaches an infrared fixed point.  There is a severe difficulty in this program, however, due
to the observed Wilson-action critical point being around $\beta = 3.2$, because the 
lattice spacing at say $\beta = 3.3$ 
is likely to be so small as to make it impossible to see the interesting region of 0.5 to 1 fm on 
practically-sized
lattices, for which the temperature will also be too high.  So one is motivated to seek 
actions that will keep the system
in the non-confining phase but allow for
a larger lattice spacing.  If one knew  what lattice artifacts were causing the transition to
confinement then an action which prohibits these objects could be constructed.  $\beta$ could then be lowered to 
increase the lattice spacing without inducing the phase transition.  For instance, in the U(1) theory, abelian
monopoles multiply as the coupling is increased eventually forming percolating chains which induces a phase
transition to a confining phase\cite{dgts}.  However, if a restriction is placed on the action 
giving such monopoles an infinite chemical potential, this theory remains in the Coulombic phase 
for all $\beta$ because the lattice
artifacts that cause confinement have been removed \cite{mitrj}.  The monopoles could also be removed
with a simple plaquette restriction, $p>0.5$, where
$p$ is the plaquette variable.  Since the continuum limit is defined in the
neighborhood of $p=1$, such a restriction does not affect the continuum limit or weak-coupling scaling of
the theory. Any objects that can be removed by a plaquette restriction $p>c$ with $c<1$ can be considered 
strong-coupling lattice artifacts which will not be operating in the continuum limit.  

The plan of this paper is to attempt the same program in the SU(2) theory, hypothesizing that confinement here
is also due to lattice artifacts which do not survive the continuum limit and can therefore be eliminated 
without affecting the continuum limit.  Whether or not the Coulomb magnetization shows spontaneous symmetry
breaking in the infinite lattice limit will be the test of whether a formulation is in the Coulombic or
confining phase.  A secondary test will be measurement of the instantaneous Coulomb potential itself 
and also the standard
interquark potential, the latter for which no gauge fixing is necessary.

We find that two artifacts must be controlled in order to prevent a transition to confinement, Z2 strings (and their
associated monopoles), and 
SO(3)-Z2 monopoles which are topologically nontrivial realizations of the non-abelian Bianchi identity. 
The latter
are gauge-invariant monopoles first introduced in \cite{so3-z2}.  To demonstrate the connection
of SO(3)-Z2 monopoles to confinement, they were monitored in standard Wilson action simulations (with no gauge fixing)
as $\beta$ was increased
in the $\beta$-region where the Coulomb magnetization transition was observed\cite{proof}.  
The monopoles  were found to form a
percolating cluster just beyond $\beta = 3.2$.  Extrapolating the percolation transition to the infinite
lattice gave a $\beta$-value precisely matching the previously identified critical point.  The monopoles 
percolate in the confining
phase. This very sharp transition may be used to locate the critical point with high precision.  Below, simulations 
which prohibit SO(3)-Z2 monopoles and also have a plaquette restriction $p>0.01$ are shown. Finite
size scaling of the Coulomb magnetization measured in configurations transformed to the minimal Coulomb gauge, shows the
system to remain
in the spontaneously broken phase on the infinite lattice, even as $\beta \rightarrow 0$.  The positive plaquette 
restriction is needed to eliminate Z2 strings, another lattice artifact known to 
be able to induce confinement.

The instantaneous Coulomb potential, which is determined from the Coulomb magnetization correlation function, 
is then measured
for this action at $\beta=0$ (the strong coupling limit), 
as is the standard interquark potential using a conventional
smeared Wilson loop approach. 
The Coulomb potential is known to be an upper limit for the interquark potential (it doesn't fully incorporate
the non-linearities 
of the gluon self-interaction in the quark color fields)\cite{zw}.
Unlike the situation for the standard Wilson action in the confining phase, where the 
Coulomb potential (and force) considerably exceeds the interquark potential (and associated force)\cite{zw,greensite}, 
here they appear to closely agree.
However, even using smeared loops, without extremely high statistics the interquark potential 
is limited by random error beyond about $R/a=20$ 
(since the force is 
smaller here than in the confining phase it is harder to measure in this system at the same lattice spacing). 
In contrast 
the Coulomb potential can be measured
out to $R/a=50$ even with relatively modest resources.  Although the Coulomb potential does not provide a
direct measurement of the force between quarks, it still provides a perfectly reasonable definition of the 
running coupling, which has the additional advantage of renormalization scheme independence\cite{zw}.
At small distances agreement with the two-loop perturbative running coupling is good, which allows 
measuring the physical lattice spacing by relating it to the $\Lambda$-parameter.  These potentials definitely
differ from those of the confining phase when scaled to equal lattice spacings.  The monopole-suppressed simulations
show a Coulombic 
behavior with running coupling $\alpha _s (R)$ at first consistent with the two-loop form but 
then slowing down, running approximately linearly up to about 1 fm, and possibly flattening
out at a value of around 1.4 at distances beyond 1.3 fm, suggestive of an infrared fixed point. 
This potential differs greatly from the linear + Coulomb form seen with the 
straight Wilson action in the  confining phase. 
 
Although the 
previous observation of spontaneous breaking of the remnant symmetry showed that both the weak coupling Wilson-action 
theory and the SO(3)-Z2 monopole-suppressed theory for all couplings were in a non-confining phase,  observation
of a Coulombic form for the potential at distances of order 1 fm shows more directly that this non-confining phase
exists and can be studied using lattice methods at hadronically interesting length scales. Because all that
has been done is elimination of lattice artifacts, this must be the phase of the continuum limit.
Therefore, one may have to look beyond gluons, to light quarks and the chirally broken vacuum for the source of confinement.

\section{Gauge invariant SO(3) monopoles}

We start with Wilson-action SU(2) lattice gauge theory with a 
positive-plaquette constraint, the positive-plaquette model\cite{pp}.
This constraint eliminates Z2 strings (strings of 
negative plaquettes).
Z2 strings are responsible for confinement in Z2 lattice gauge theory,
so eliminating them causes the Z2 theory to deconfine.
The positive-plaquette SU(2) model, however, 
still confines at small $\beta$ \cite{hellerpp}. 
So in SU(2) there must be something besides Z2 strings that causes
confinement. Actually Z2 strings probably {\em are}
responsible for 
confinement in the mixed fundamental-adjoint \cite{bc} version of
SU(2) in the large $\beta _A$ region, which includes the
Z2 theory as a limiting case. Because Z2 strings can cause confinement
(though they are not the only cause) a positive plaquette constraint
must be maintained along with any monopole constraint 
in order to get a possibly non-confined theory. In practice we modify this
constraint to $p>0.01$, because although $p>0$ appears to work,
some instability leading to larger statistical errors is seen for that case,
perhaps because one is right on the edge of a transition. (A short run
with a $p>-0.1$ restriction together with the monopole restriction detailed below
showed a definite lack of remnant symmetry breaking,
signaling confinement).

The identification of the monopole starts with
the non-abelian Bianchi identity\cite{bal,skala}.
\begin{figure}[ht]
                      \includegraphics[width=3.25in]{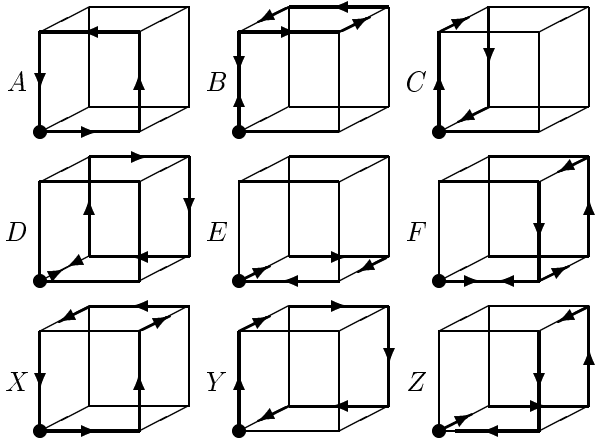}
                                  \caption{Covariant plaquettes and bent double-plaquettes 
for the lattice non-abelian Bianchi identity.}
          \label{fig1}
       \end{figure}
This can be expressed by first constructing 
the covariant (untraced) plaquettes
that comprise the six faces of an elementary cube, with the necessary
``tails'' to bring them to the same starting site (Fig.~1). Call these
$A$, $B$, $C$, $D$, $E$, and $F$. Now construct three bent double 
plaquettes also shown in Fig.~1, $X=AB$, $Y=CD$, and $Z=EF$.  
If one forms the
product $XYZ$, each link will exactly cancel with its conjugate, so
$XYZ=\mathds{1}$, the unit matrix. This is the non-abelian Bianchi identity. 
Although the plaquettes
are all positive (due to the positive plaquette
constraint), and thus have a trivial Z2 component of unity, the
double plaquettes may be negative. Factor each of 
these into  Z2 and SO(3) (positive-trace) factors, e.g. 
$X=Z_{X}X'$ etc.,
where $Z_{X}=\pm 1$ and $\mathrm{tr} X' > 0$.  Then the Bianchi identity
reads $X' Y' Z' Z_{X} Z_{Y} Z_{Z}=\mathds{1}$. This can be realized in either 
a topologically
trivial or nontrivial way as far as the SO(3) group is concerned.
If $Z_{X} Z_{Y} Z_{Z}=1$ then $X'Y'Z'=\mathds{1}$. However if 
$Z_{X} Z_{Y} Z_{Z}=-1$ then
$X'Y'Z'=-\mathds{1}$. In this case one has an SO(3) monopole which,
since it also carries a Z2 charge, can be pictured to be at the same
time a Z2 monopole. Both possible operator orderings need to be checked.
The decomposition of the double-plaquettes
into Z2 and SO(3) factors is gauge invariant
since their traces are invariant. 
In such a monopole a large SO(3) flux is
in some sense cancelled by a large Z2 
flux in order to satisfy the SU(2)
Bianchi identity.  This is reminiscent of the abelian monopole
in U(1), in which a large net flux of $2\pi$ enters or 
exits an elementary cube.
This apparent non-conservation of flux 
is allowed by the compact Bianchi 
identity since $\exp (2\pi i)=1$.
In the continuum the Bianchi identity enforces 
exact flux conservation.
If plaquettes in U(1) are restricted 
to $\cos(\theta _p) > 0.5$, then
the only solution to the 
Bianchi identity is the topologically trivial one,
$\theta_{\mathrm{tot}}=0$,
where $\theta _{\mathrm{tot}}$ is the sum 
of the six plaquette angles in an
elementary cube. This eliminates the monopoles,
and shows that they 
are strong-coupling lattice artifacts. 
The U(1) lattice gauge theory,
as a result, is deconfined in the continuum limit.

The SO(3) monopoles described above 
are also lattice artifacts.
If plaquettes are restricted 
so that $\cos (\theta _{p}) \! > \! \sqrt{2}/2$, 
then even the double-plaquettes are positive,
and the SO(3) monopoles described above cannot exist.  Since in
the continuum limit all plaquettes are in the neighborhood 
of the identity,
such a restriction should have no effect on the continuum limit.
Therefore, these monopoles will not exist in the continuum, and 
exact SO(3) flux conservation on elementary cubes will hold there.  
Indeed it
has long been recognized that if SU(2) confinement is due 
to monopoles or vortices, then the only
such objects which could survive the 
continuum limit to produce 
confinement there are large objects (fat monopoles
and vortices) for which flux
is built up gradually\cite{fat}.
A good way to look
for fat-monopole confining configurations would 
seem to be 
to choose an action
which eliminates the single lattice spacing
scale artifacts while still allowing
similar larger objects to exist. 

The plaquette constraint mentioned above is one such possibility, 
however this results in a rather weak renormalized coupling as does a
positive constraint on all bent double plaquettes.  
A weak renormalized coupling results in a small lattice spacing 
which prevents one from studying hadronically interesting scales
and low enough physical temperatures on lattices of
a practical size.  The best solution, if one
wants to eliminate all monopoles is to put in an infinite
chemical potential for monopoles themselves.  This is done simply
by rejecting any update that creates a monopole.  We find that
such a restriction, along with $p>0.01$ and with $\beta =0$, results
in a lattice spacing, as determined by fits of the running coupling to
the two-loop perturbative form, approximately equal to that of the standard Wilson
theory at $\beta = 2.85$.  This allows our $40^4$ and $60^4$ lattices to probe
the usual region of interest for lattice potentials, 0.2-1.5fm, with
physical temperatures well below the usual deconfinement temperature.
One could obtain even larger lattice spacings by backing off the 
chemical potential from infinity and allowing some monopoles. 
So long as they do not percolate they are not expected to cause a phase transition.
However, they are still powerful lattice artifacts which could affect detailed
numerical results, so it would seem most prudent to eliminate them all
if possible,
which with today's technology, even on ordinary PC's, is practical.

In order to demonstrate their possible connection to confinement, 
these monopoles were previously studied in the standard Wilson theory, without gauge fixing.
These simulations used a standard heat-bath alternated with overrelaxation algorithm, and periodic
boundary conditions.
The region around $\beta=3.2$ was studied on various lattices from $16^4$ to $40^4$\cite{proof}, as the 
Coulomb gauge study\cite{coulgauge} had seen a zero-temperature 
deconfinement transition extrapolated to the infinite lattice at $\beta =3.18\pm 0.08$.
These monopoles do not form closed loops on the dual lattice, because there is
no exact conservation law that would force this, however they do cluster, and one can still
search for a percolating cluster.   If one takes the 50\% percolating level as
defining the finite lattice percolation point then one can extrapolate to the infinite
lattice which gives an infinite lattice percolation point of $3.19 \pm 0.03$\cite{proof}.  The sharpness
of the percolation probability curves gives a very high precision.  Most of the uncertainty comes
from the infinite lattice extrapolation. 
The agreement of this percolation threshold with the previously determined critical point from the 
Coulomb gauge magnetization
gives credence to the idea that these monopoles could be responsible for confinement.  It
also supports the previous determination in that the percolation study used neither Coulomb
gauge fixing nor the unconventional open boundary conditions of the previous study. 

The SO(3)-Z2 monopoles are ubiquitous even at these relatively weak couplings, occupying approximately
13\% of the dual lattice links at the percolation threshold.  If these artifacts strongly
affect other measured quantities, which is possible, then even in the deconfined region for 
$\beta > 3.2$ the Wilson action may give poor results. 

 The next demonstration of the possible connection
of these monopoles to confinement involves applying the monopole constraint suggested above (preventing
any monopoles from forming) along with the plaquette constraint $p>0.01$.  In order to give these lattices
a maximum chance to confine, and as large as possible lattice spacing, 
the simulations are performed in the strong coupling limit, $\beta=0$.  
An 8-hit Metropolis algorithm is used. Since the constraint is an accept/reject decision, heat bath and Metropolis are
almost the same for the $\beta=0$ action, with the difference that the Metropolis gives up after a certain number of 
update attempts. After each sweep, the Coulomb Gauge is set using an overrelaxation algorithm, 
with an overshoot of 0.7.  
On the lattice, one attempts to set the minimal Coulomb gauge, where one maximizes the sum of traces of
all spacelike links.
One serious problem with this procedure, which has limited its usefulness,
is that different minimization runs (if preceded by a 
random gauge transformation) find different local maxima for which measured quantities may differ
substantially (typically $\pm 4$\% for the average magnetization of the 4th dimension pointing links).  
This ``lattice Gribov problem" has
made obtaining precision results in the Coulomb gauge very difficult.
In Ref.~\cite{coulgauge}, however,
it was shown that this difficulty is almost eliminated by using open boundary conditions, which remove the
global constraints imposed by the gauge-invariant Polyakov loops which appear to be causing the local overrelaxation
algorithm to get hung up on local maxima. With open boundary conditions, variations between different maximizations
are {\em several hundred} times smaller than with periodic boundary conditions, making the uncertainty introduced by the Coulomb
gauge fixing comparable to or smaller than the random errors of the simulation for the simulation lengths shown here.
For a similar reason in a recent study of topological charge, 
L\"{u}scher and Schaefer have also found open boundary conditions to be useful, and they have justified their use in gauge
theories\cite{luscher}.  The Coulomb magnetization $<|\vec{m}|>$, is defined from 
\begin{equation}
\vec{m}_i = \frac{1}{L^3}\sum _{\mathrm{hyperlayer}} \vec{a}
\end{equation}
which is the magnetization on the $i^\mathrm{th}$ hyperlayer. The expectation value above includes a sum over hyperlayers
as well as configurations.
Here $\vec{a}=(a_0$,$a_1$,$a_2$,$a_3 )$ is the O(4) color-vector associated with the fourth-direction pointing gauge element,
\begin{equation} 
U_{\vec{r},4}=a_0 \mathds{1} + \sum _{j=1} ^3 ia_j \tau _j ,
\end{equation}
the $\tau _j$ being Pauli matrices.
The Coulomb magnetization and associated
Binder cumulant, $U=1-\! <\! |\vec{m}|^4 \! >/(3\! <\! |\vec{m}|^2 \! >^2 )$, are measured on various lattices.
From 10,000 to 50,000 sweeps are averaged, after 5000 equilibration sweeps (10,000 for the $40^4$ and $60^4$ lattices). 
Quantities were tracked during
equilibration to be sure it was sufficient. Trial new links for the Metropolis algorithm were taken
from the nearest one-half of the possible SU(2) gauge space surrounding the link.  This resulted
in an acceptance rate of 14\%. So it takes about four of these sweeps to equal one ordinary
sweep with a 50\% acceptance.  Later tuning of the algorithm showed that restricting the update matrix
to $\mathrm{tr}(U_{\mathrm{update}})>0.5$, with an acceptance rate of $\sim 30$\% would maximize the algorithm's speed through configuration space.
A $50\%$ acceptance is
overall slower because rejections are on average quicker than acceptances.  This is because as soon as a
monopole is detected, the update can be rejected without checking the remainder of the neighborhood.
Runs with a completely open update were also performed to check ergodicity.  These results agreed with the others.

The region of
the lattice six or fewer lattice spacings from the open boundary is excluded from measurements.  The remaining
boundary effects on measured quantities were of order the random errors.  No matter where this cut is made, the 
remaining boundary effects can be studied through finite size scaling. However, some exclusion of the 
boundary seems necessary to obtain easily interpretable results on reasonable sized lattices.
The exclusion of the boundary region
can be considered a kind of ``soft" or ``dynamical" open boundary condition on the interior lattice.  Some data 
were also collected 
with a smaller exclusion zone of four lattice spacings.  These were rather similar, but with somewhat larger
differences between on-axis and body-diagonal potentials.

\begin{figure}[ht]
                      \includegraphics[width=3.25in]{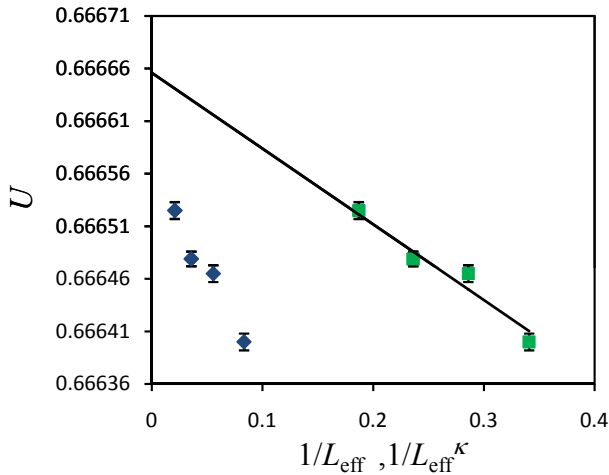}
                                  \caption{Binder cumulant for monopole-suppressed simulations with various lattice sizes plotted
against two different abscissas. Points on left are vs. $1/L_{\mathrm{eff}}$.}
          \label{fig2}
       \end{figure}
For the O(4) order parameter (Coulomb Magnetization) the Binder
cumulant is expected to approach 1/2 in an unmagnetized phase as the lattice size becomes infinite and 2/3 in
a magnetized (spontaneously broken) phase\cite{mdop}.  In Fig.~2, $U$ is shown vs. $1/L_\mathrm{eff}$ where 
$L_\mathrm{eff} =L-12$ and $L$ is the linear lattice size 
which ranged from 24 to 60. $L_\mathrm{eff}$ is the size of the region inside the exclusion zone.
One sees that $U$ is very close to 2/3 already, even for the smallest lattice size
and is increasing with lattice size. This is the expected behavior in a symmetry-broken phase.
To explore the extrapolation further, $U$ was assumed to
behave as $1/L^\kappa$, where $\kappa$ is an unknown constant.  A value of $\kappa=0.4$ gave the best fit.  The second plot shows 
$U$ vs. $1/L^k$, showing a consistent extrapolation to $2/3$ as $L \rightarrow \infty$.  Since one is within a phase here
and not necessarily near a critical point, but also not deep within the phase (susceptibility is still increasing somewhat with lattice size),
the expected scaling is not predicted from finite-size scaling theory.  The evolution of $U$ with lattice size depends on
the scaling function which is not universal, so the best one can do is a phenomenological fit.
Fig.~3 shows that, also consistent with a spontaneously broken phase, the Coulomb magnetization is 
extrapolating to a nonzero value in the infinite lattice limit.  The axes were chosen such that at a critical point
one would expect a straight line through the origin. A spontaneously broken phase lies above this line and a symmetric
phase lies below it.
\begin{figure}[ht]
                      \includegraphics[width=3.25in]{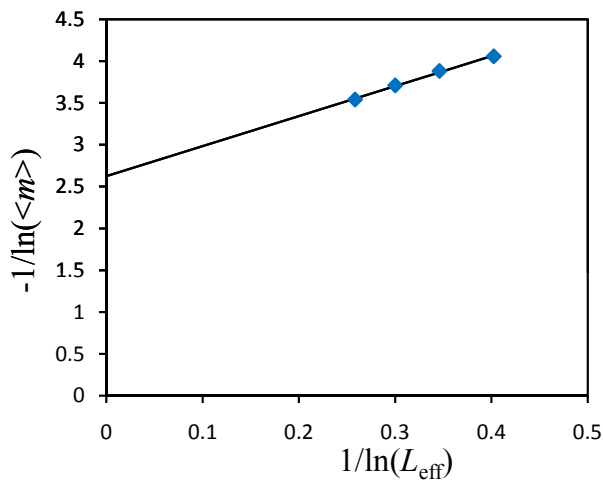}
                                  \caption{Coulomb magnetization for monopole-suppressed simulations with various lattice sizes.}
          \label{fig3}
       \end{figure}
If the remnant symmetry is broken at $\beta =0$, it almost certainly will remain broken at finite $\beta$, where the
configurations are more ordered, therefore the entire $\beta$ range in the monopole-suppressed theory including
the continuum limit would appear to be in a phase of broken remnant symmetry.  It has been shown that such
a phase is necessarily non-confining, because if the magnetization is nonzero, the lattice Coulomb 
potential (defined below)
must approach a constant at long distances\cite{zw}. Because the lattice Coulomb potential is an upper limit to the 
interquark potential, the latter also is prohibited from having a linear term in the symmetry-broken phase. 
Supporting the above, a different method
for removing violations of the non-abelian Bianchi identity also appears to remove confinement in 
four dimensions (but not three)\cite{nabi}.

Thus, it appears that eliminating SO(3) monopoles (and also Z2 strings from the plaquette constraint) eliminates
confinement from the 4-d SU(2) gauge theory. However, it is important to show that the lattice spacing for this formulation is large enough 
for physically interesting length scales to be probed, and also so
that the physical temperature associated with our largest lattice is in the region where confinement 
would be expected.  This we will do through measurement of the lattice Coulomb potential, and from that, the running coupling.

\section{Lattice Coulomb Potential}

   The lattice Coulomb potential, $V(R)$ is given as a function of the Coulomb magnetization equal-time two point
correlation function\cite{zw}, 
\begin{equation}
aV(R)=-\mathrm{ln}(<\vec{a}(\vec{r}_1 )\cdot \vec{a}(\vec{r}_2 )>) .
\end{equation}
Where the two links are on the same spacelike hyperlayer with separation $R= | \vec{r}_2 -\vec{r}_1 |$.
The expectation value is over both configurations and hyperlayers, avoiding the six closest to either boundary.
It differs from the interquark potential in that the latter is derived
from Wilson loops with a long time extent which allows the string connecting the quarks to attain its lowest energy
state.  The Coulomb potential, by contrast, creates the quark-antiquark pair with associated gluon field
for an instant, so nonlinear effects of the response of the gluon vacuum are not fully included.  However, as mentioned
before, the Coulomb potential is an upper limit for the interquark potential, so it contains very useful information. 
In addition, the Coulomb potential gives a perfectly reasonable definition of the running coupling, a quantity,
the behavior of which would be very interesting to know outside the perturbative region, and which within the
perturbative region can also 
be used to determine the lattice spacing.  Because it is determined from a simple link correlation function as opposed
to Wilson loops large in two dimensions, 
the Coulomb
potential can be measured to a considerably larger distance than the interquark potential, which, even when using 
smeared operators,
is rather quickly swamped by random errors.  This makes the Coulomb potential a very exciting quantity to work with, as
it opens the possibility of clearly seeing the infrared behavior of the running coupling.  Another considerable advantage of the 
Coulomb potential is that it is directly calculated - no fitting is required.

The correlation function is measured both on-axis (OA) and along body diagonal (BD) direction lines on lattices up to $60^4$.  One can 
determine the effects of the finite lattice on the potential, force, and running coupling by comparing the 
different lattice sizes. 
The running coupling is defined from the potential through
\begin{equation}
\alpha(\sqrt{R_1 R_2})=\frac{4}{3}R_1 R_2 (V(R_2 )-V(R_1 ))/(R_2 - R_1 )
\end{equation}
where $R_2$ and $R_1$ are distances from an initial point to adjacent lattice sites, 
either on-axis or along a body diagonal.
\begin{figure}[ht]
                      \includegraphics[width=3.25in]{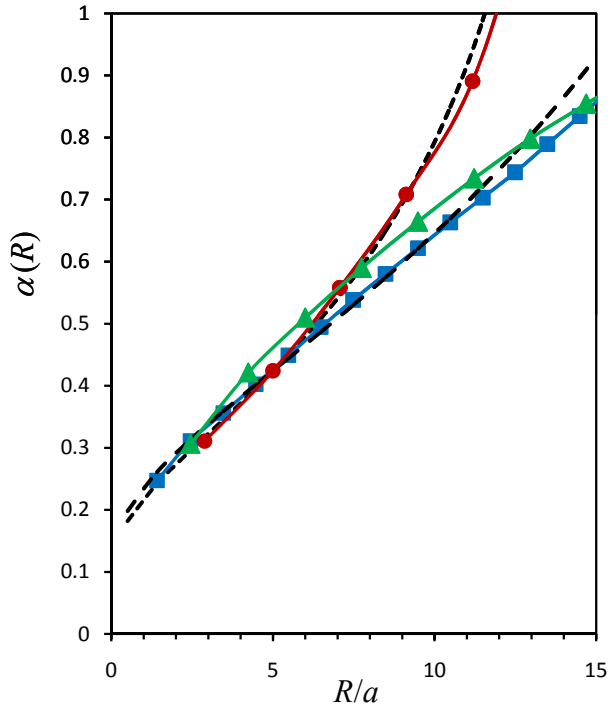}
                                  \caption{Running coupling in the low-R region for the $60^4$ lattice.
Squares are OA, triangles BD. Circles are UKQCD collaboration Wilson-action data. Short-dashed line is fit to two-loop 
perturbative form, long-dashed line is one-loop.  Error bars are smaller than plotted points.}
          \label{fig4}
       \end{figure}
Fig.~4 shows the running coupling for the $60^4$ lattice 
in the small-R region.
Some degree of rotational non-invariance is seen between the OA and BD results, which seems
to affect small-distance and large-distance results differently.  More sophisticated averages of the 
two $R$ values used in the force determination based on the free-field lattice Fourier transform were tried, but did not
reduce the rotational non-invariance, so the simpler geometric average was used, as shown above. The free field may not be a 
good guide to the interacting case.
Also plotted is the
one-parameter fit of the OA data to the two-loop 
renormalization group improved perturbation theory form\cite{creutz}
\begin{equation}
\alpha (R) = \left( 4\pi b_0 \left[\ln ((R_0/R)^2)
+(b_1/b_{0}^{2})\ln \ln((R_0/R)^2)\right] \right) ^{-1} 
\end{equation}
(where $b_0 =11/24\pi^2$ and $b_1/b_0^2=102/121$)
in the range $R=2a$ to $R=6a$. The OA data gives a smaller lattice spacing, so that is the more conservative
choice, although averaging the lattice spacing obtained from
the two datasets would probably make more sense, since they are likely extremes of the rotational non-invariance.
Also shown is Wilson-action OA data of the UKQCD collaboration\cite{ukqcd} for $\beta=2.85$, with the lattice spacing scaled for
best fit at $R=5a$, which gives a factor of 0.98, the $\beta=2.85$ lattice spacing 
being slightly larger than that of
the monopole-suppressed
simulation (but the same within errors).  Our $60^4$ lattice is therefore physically slightly larger than 
that of the UKQCD simulation ($48^3 \times 56$), 
so there is little worry
that the lattice is too small to access a region where confinement should easily be seen, if there.  Also the 
temperature for a lattice of this size is about 1/2 of the finite-temperature deconfinement temperature, so our
lattices could not be deconfined for this reason.  The fit to the running coupling gives $R_0 /a =23.5$. In this
renormalization scheme $1/R_0$ is very close to $\Lambda _{\overline{\mathrm{MS}}}$.  Since we are working with SU(2) and
not the physical SU(3), it is not worth setting the scale to a high degree of precision.  Taking $R_0^{-1}$=200 MeV,
gives $a=0.043$fm  with a largest lattice dimension of $60a=2.6$fm.

\begin{figure}[ht]
                      \includegraphics[width=3.25in]{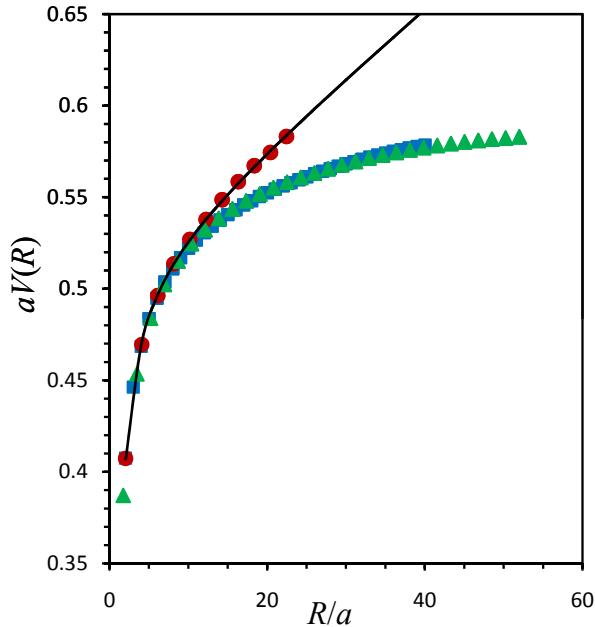}
                                  \caption{Coulomb potential for on-axis (squares) and body-diagonal(triangles).  Also shown is
UKQCD data for the interquark potential (circles). Error bars are smaller than plotted points. Fit shown to UKQCD data is linear+Coulomb.}
          \label{fig5}
       \end{figure}
\begin{figure}[ht]
                      \includegraphics[width=3.25in]{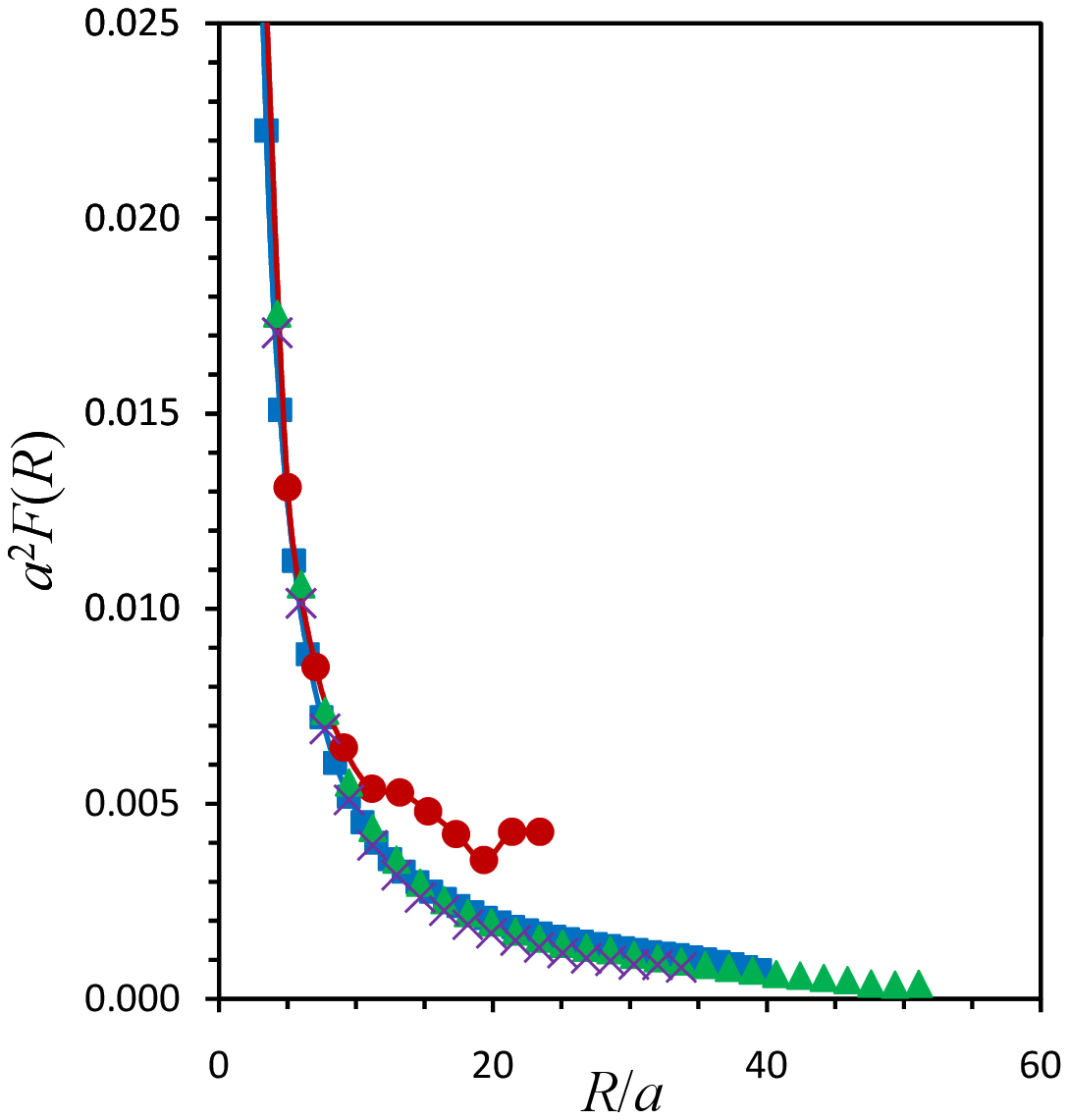}
                                  \caption{Coulomb force calculated through finite differences from the Coulomb potential for
on-axis(squares) and body diagonal (triangles). Also shown are the $40^4$ BD data ($\times$), and UKQCD interquark force (circles).
Errors are smaller than plotted points.}
          \label{fig6}
       \end{figure}
Now that the relative lattice spacings for the monopole-suppressed and Wilson-action simulations have been determined,
the respective potentials (Fig.~5) and forces (Fig.~6) can be compared.
For the potential a constant of 0.061 must be added to shift the UKQCD results to match at $R=4a$, and 0.0075 is
added to the body-diagonal values for the same purpose.
For potentials defined differently a different additive renormalization is expected.  This has no effect on the physical forces. 
Although the two formulations
agree in the perturbative small-R region, the monopole-suppressed theory clearly shows a more Coulombic form which
contrasts strongly to the linear large-R confining potential of the $\beta=2.85$ Wilson-action data.  The situation 
is most clear from the force graph.  Whereas the UKQCD data show clear evidence of the force approaching a
non-zero constant at large distances (string tension),  no such trend is visible in the monopole-suppressed data,
for which the force appears to be vanishing at large distances.
The contrast is striking - clearly the two
simulations cannot be made to agree.  Differences between OA and BD and between $40^4$ and $60^4$ are
present, but appear to be relatively minor.

\begin{figure}[ht]
                      \includegraphics[width=3.25in]{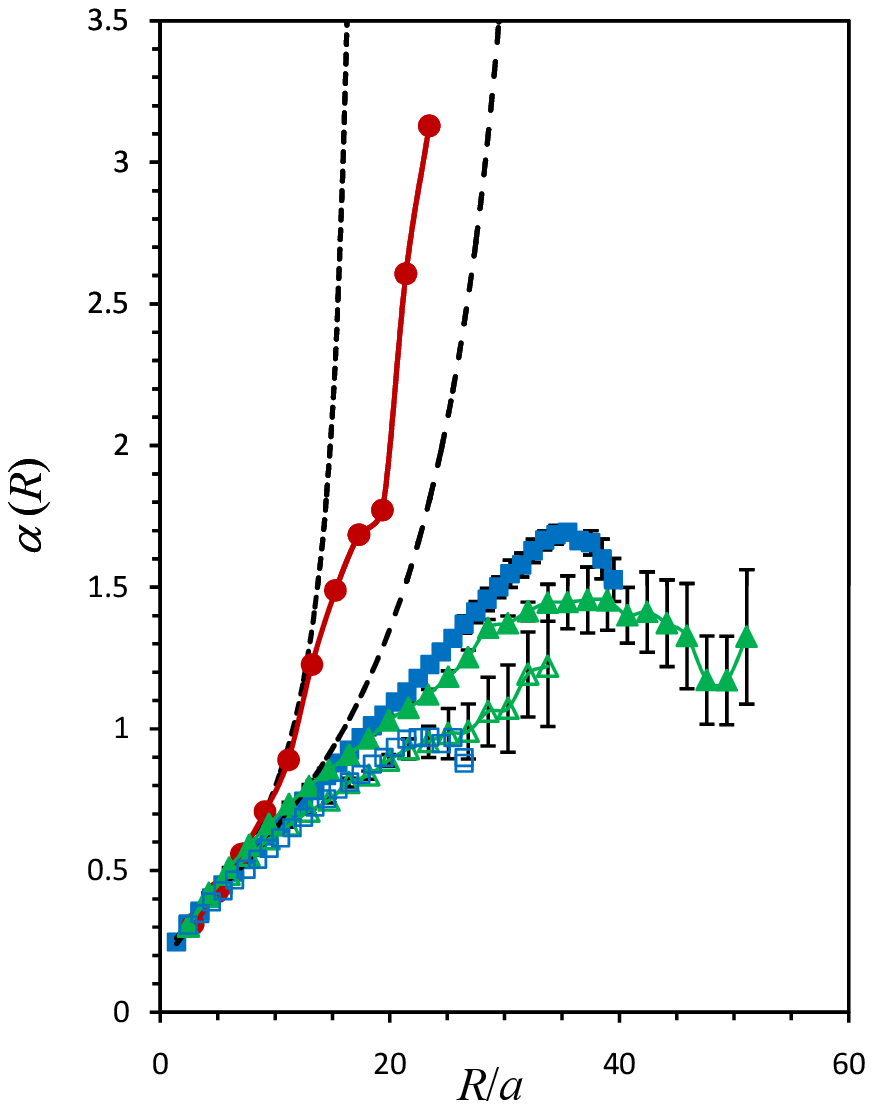}
                                  \caption{Running coupling for large-$R$ for $60^4$ OA (squares), BD (triangles) and the same for
$40^4$ (open symbols), along with the running coupling calculated from the UKQCD interquark force (circles). Error bars are from binned
fluctuations.  Adjacent errors are highly correlated, so taking two-step finite differences does not reduce the error significantly.
Short dash line is two-loop fit from Fig.~4 and long dashed is one-loop.  The two-loop Landau pole is at $R=19a$.}
          \label{fig7}
       \end{figure}
Finally, the running coupling, defined above, is shown for the full lattice (Fig.~7).  Here the differences between 
OA and BD and between $40^4$ and $60^4$ can be examined in more detail. Whereas the small-R differences
between the OA and BD results are likely due to finite spacing rotational non-invariance,
the large-R differences are more likely a diverging response of the different objects to finite lattice size effects.
The OA and BD running couplings agree within better than 10\% and follow similar trends until $R=0.5L$ for both $40^4$
and $60^4$.  Because
of the open boundary condition,  separations beyond $R=0.5L$ can be considered since there is no second path through
the boundary.  However, beyond $R=0.5L$ the BD and OA no longer agree, so one can assume that the OA is feeling the
finite size effect more severely here.  One would expect that the BD would be reliable out to a separation
$\sqrt{3}$ larger than the OA simply because there is that much more extension available along the body diagonal.  
This is borne out by the $40^4$ data which  roughly follows the $60^4$ trend, though lies below it by 10-20\%,
out to $20\sqrt{3}a=35a$.  This would suggest that the $60^4$ BD data should be reliable at this level out 
to $30\sqrt{3}a=52a$, which is as far as points are plotted in the figure.  
Up until $R=20a$ $\alpha(R)$ shows a roughly linear trend, but is slightly concave downward
(as also is the one and two-loop result initially).
Beyond $R=20a$ the OA potential straightens, whereas the BD continues concave downward, flattening beyond
$R=35$.  This is highly suggestive of an infrared fixed point.  However,  one cannot be absolutely sure that
the BD potential is reliable beyond the place where it leaves the OA potential $R=30a$,  so we are reluctant
to claim an infrared fixed point without also seeing the signal in an OA potential. This, unfortunately, would
require a lattice of $72^4$ or larger, which is beyond our present capacity.  
Simulations on an asymmetric $60^3\times 75$ lattice are currently being run in an attempt to access the OA potential
out to $R=37a$.
One additional point in favor of an infrared fixed point is that if 
one determines the second derivative of $\alpha$ in the region $R<10a$ where it can be reliably determined due
to small random errors here, and extrapolate where the slope would vanish if this trend continued, that also
yields a point of maximum $\alpha$ around $R=40a$. It is worth noting that $\alpha (R)$ does not necessarily
have to have a fixed point in order to obtain a non-confining potential. Indeed it 
can actually still diverge
by any power less than unity, and the potential will still be non-confining (become constant at $R\rightarrow \infty $).  Thus
the behavior seen here for $\alpha$ which either shows an infrared fixed point, or possibly still
diverging with a slightly under linear (concave-down) behavior is fully consistent with a non-confining potential.

\section{Interquark Potential}
  In the last section, we were comparing two somewhat different potentials, the interquark potential for the 
Wilson-action data and the Coulomb potential as defined in the minimal Coulomb gauge, which is expected to be
an upper limit for the interquark potential.  About one PC-year of computing time was devoted to the above study. 
At least two PC-years were devoted to determining the interquark potential using a more standard smeared operator method\cite{ukqcd,smear}.  
The latter simulations
were done on a $40^4$ lattice with ordinary periodic boundary conditions and no gauge fixing.  A total of 200,000 sweeps were
performed, with 5000 discarded for equilibration and loops measured every 50 sweeps.
Smeared Wilson loops to size 19x19 were measured.  Three different smearing levels (5, 10 and 20 iterations)
were generated using the recursive blocking scheme which replaces links with a combination of the original link
and U-bends.  A straight-link weight of $c=2$ as defined in $\cite{ukqcd}$ was used.
Wilson loops with both like and unlike operators at the ends were measured resulting 
in six different types of loops. As usual, timelike links were not smeared in order to retain the transfer-matrix
interpretation.  Some larger smearing levels were tried, but an indication that the simulations 
were possibly becoming 
sensitive to the lattice size through the smearing operation led us to cut the number of smearings so that 
information on the finite lattice size would not be fed to the observables.  For each $R$, the $6\times 18$ ($T=1$
excluded)
smeared Wilson loops, $W_{ik}(R,T)$ were fit to a triple exponential form
\begin{equation}
\sum_{j=1}^{3}p_{ij}p_{kj}\exp (-V_j (R)T)
\end{equation}
where $p_{ij}$ are overlap coefficients between the given smeared operator ($i=1$..3) and the
energy eigenstate ($j=1$..3).  Since no secular trend was obvious in the excited state energies $V_2$ and $V_3$
as $R$ was varied, fits were then redone using fixed average values for the excited 
state energies of 0.72 and 1.33 in lattice units.  
This resulted in 
lower errors on $V(R)=V_{1}(R)$ while still giving reasonable $\chi ^2$/d.f. values 
averaging 1.7.  Errors in the smeared loops themselves were determined from binned fluctuations. 
We only studied the interquark potential on axis,
because it was clear that our statistics would not allow measurement much beyond $R=18a$ which was already
quite challenging.  
\begin{figure}[ht]
                      \includegraphics[width=3.25in]{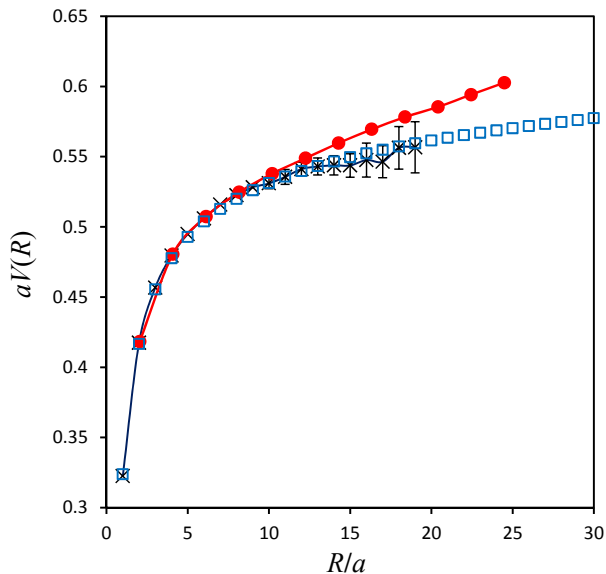}
                                  \caption{Interquark potential from smeared loops ($\times$) together with
the Coulomb potential from Coulomb magnetization described above (squares) and UKQCD interquark potential (circles), scaled
for slight difference in lattice spacing.}
          \label{fig8}
       \end{figure}

In Fig.~8 the interquark potential 
and the OA Coulomb potential are shown together, along with the UKQCD potential. 
These are all adjusted with an additive renormalization constant to agree at $R=4a$. 
The Coulomb and interquark potentials for the monopole-suppressed case appear to agree.
The idea that the Coulomb potential is an upper limit of the interquark potential is really only
a statement on long distance behavior, because the two differ by a finite additive normalization
which is not necessarily larger for the Coulomb case.  The monopole-suppressed interquark potential
also appears to fall below the UKQCD result, agreeing with the trend seen earlier with the Coulomb potential
which is determined to a much higher precision and as a result, to a longer distance.
Agreement between the interquark and Coulomb potentials
suggests that the quark color field produced by the link operators in the Coulomb gauge-fixed
configuration may be quite close to the physical fields. This supports the observations of Ref.~\cite{coulombic} where
it was shown that a superposition of single-quark Coulomb fields 
gives a more accurate depiction 
of the ground state color fields of the two-quark system than a narrow flux tube does. 
This may mean that 
the effects of nonlinearity in the SU(2) color field are not actually that large.

Use of open boundary conditions with Coulomb gauge fixing is 
a new technology, so it is important to see that more standard methods using periodic boundary conditions
and without any gauge fixing yield similar results.  The precision of the Coulomb potential defined
in the Coulomb gauge makes the extra effort of gauge fixing more than worthwhile.  Because the force is taken
from the change in potential from one $R$ to the next, which is very small, the potential must be measured to
a very high accuracy in order to measure $\alpha (R)$.  At $R=12$ the interquark potential 
simulation has an error of 0.9\%.  The more heavily smeared UKQCD simulation which had similar statistics achieved
a random error of 0.16\%.  The corresponding error in the Coulomb potential is only 0.06\%, with about $1/6$ the effort.
Plus there is no (systematic) uncertainty from the fitting procedure. Of course one needs to remember that these are different quantities, but 
for this system they appear to very similar, and both can be used to obtain the running coupling.
The error in potential is
approximately linear in $R$.  With force going like $1/R^2$ at the larger $R$'s its relative error grows as $R^3$.  
With the current statistics (a four month PC run) the random error on the force (and alpha)
is 5\% at $R=34a$.  With supercomputing, combined with better tuning of the algorithms, one could imagine measuring Coulomb potentials 
out to distances of 80-100 lattice spacings using these techniques.

\section{Conclusion}

In this paper SO(3)-Z2 monopoles, which result from a topologically nontrivial realization of the non-abelian
Bianchi identity are completely suppressed, along with Z2 strings and monopoles.
Simulations were performed at $\beta=0$, the strong coupling limit, so the action consisted entirely 
of these two constraints.  Since all plaquettes are in the neighborhood of the identity in the continuum limit,
these constraints should not affect it.  This action, in the strong coupling limit, gives a short-distance
potential similar to the Wilson action at $\beta=2.85$.  A fit to the two-loop running coupling allows one to
find the physical lattice spacing, showing that the largest ($60^4$) lattice is (2.6fm)$^4$.  The interior region of
this open boundary condition lattice where measurements were made ($48^4$) is (2.1fm)$^4$.  Here we are using physical
scales that strictly apply only to SU(3) also to the SU(2) case as is usual. These lattices therefore probe the
region of interest for quarkonium states, where Wilson-action simulations see a linear+Coulomb potential. The
potential seen with the monopole suppressed action is much more Coulombic, and can entirely be fit with a 
Coulomb potential with a moderately running potential, one which is flattening out at the largest distances 
measured, possibly indicating an infrared fixed point of around $\alpha=1.4$.  There is no evidence for a linear
term which is entirely consistent with the observation of spontaneous breaking of the Coulomb-gauge remnant symmetry.
Such symmetry breaking precludes the existence of linear confinement.  
Making $\beta$ larger will only order the configurations more, so there is almost no chance this ordered 
symmetry breaking
would not continue to hold for $\beta >0$, including the continuum limit, $\beta \rightarrow \infty$.
These results show a clear lack
of universality with the Wilson action.  Since all that has been done is to remove strong-coupling artifacts,
the conclusion one is led to is that the confinement seen with the Wilson action is due to these strong-coupling
artifacts, similar to the U(1) case, and a deconfining phase transition exists in the zero-temperature Wilson
action case.  This is contrary to the usual lore, but such a phase transition has be seen in the Coulomb-gauge
magnetization at $\beta _c = 3.18(8)$\cite{coulgauge}, which is consistent with the SO(3)-Z2 monopole percolation transition
at $\beta = 3.19(3)$ \cite{proof}. 

In the intermediate distance region of $R=5a$ to $R=28a$ (0.2-1.2fm), the running coupling increases roughly linearly, which means
that the Coulomb force here is more like $1/R$ than $1/R^2$. If one integrates to get an effective potential for this region
one obtains a logarithmic potential.  Interestingly, phenomenological fits to quarkonium with logarithmic potentials\cite{logpot}
or very small powers of $R$\cite{martin}
work rather well. One may not need a linear term in the potential to explain these systems.
When light quarks are added to the theory, absolute confinement is destroyed anyway, so that may not be
a necessary ingredient of a successful theory of the strong interactions.  Sparking of the vacuum\cite{gribov}
and/or rearrangement of the chiral condensate\cite{ccideas,cahill} may itself prevent color non-singlet states from existing.

\section{Acknowledgement}
Phillip Arndt helped with some of the computations for this paper.

\end{document}